\documentclass[letterpaper, 10 pt, conference]{ieeeconf}  

\IEEEoverridecommandlockouts                              

\usepackage{listings}
\usepackage{algpseudocode}
\usepackage{algorithmicx}
\usepackage{algorithm}
\usepackage{lscape}

\usepackage{import}
\usepackage{epsf}
\usepackage[pdftex]{graphicx}
\usepackage{epsfig}

\usepackage{latexsym,amssymb}
\usepackage[table]{xcolor}
\usepackage{tikz}
\usetikzlibrary{shapes.geometric,arrows}
\usepackage{pgfplots}

\usetikzlibrary{arrows,shapes}
\usepackage{verbatim} 
\usepackage{comment}
\usepackage{wrapfig}
\usepackage{alltt}

\usepackage{color, xcolor, colortbl}
\usepackage{multirow}
\usepackage{subfig}
\usepackage[OT1]{fontenc}
\usepackage{anysize}
\usepackage{listings} 

\usepackage[style=altlist,toc,numberline,acronym]{glossaries}
\usepackage{url}
\usepackage{amsmath}
\usepackage{makeidx} 
\makeindex  

\title{\LARGE \bf
One-Shot Speaker Identification for a Service Robot using a CNN-based Generic Verifier
}

\author{Ivette V\'elez$^{1}$, Caleb Rascon$^{2}$ and Gibr\'an Fuentes-Pineda$^{3}$
\thanks{Instituto de Investigaciones en Matem\'aticas Aplicadas y en Sistemas (IIMAS), Universidad Nacional Aut\'onoma de M\'exico (UNAM), Mexico.}
\thanks{$^{1}$ {\tt\small ijvelezt@gmail.com}}%
\thanks{$^{2}$ {\tt\small caleb.rascon@iimas.unam.mx}}%
\thanks{$^{3}$ {\tt\small gibranfp@unam.mx}}%
}

\begin{document}

\maketitle
\thispagestyle{empty}
\pagestyle{empty}

\begin{abstract}
In service robotics, there is an interest to identify the user by voice alone. However, in application scenarios where a service robot acts as a waiter or a store clerk, new users are expected to enter the environment frequently. Typically, speaker identification models need to be retrained when this occurs, which can take an impractical amount of time. In this paper, a new approach for speaker identification through verification has been developed using a Siamese Convolutional Neural Network architecture (SCNN), where it learns to generically verify if two audio signals are from the same speaker. By having an external database of recorded audio of the users, identification is carried out by verifying the speech input with each of its entries. If new users are encountered, it is only required to add their recorded audio to the external database to be able to be identified, without retraining. The system was evaluated in four different aspects: the performance of the verifier, the performance of the system as a classifier using clean audio, its speed, and its accuracy in real-life settings. Its performance in conjunction with its one-shot-learning capabilities, makes the proposed system a viable alternative for speaker identification for service robots.
\end{abstract}

\section{INTRODUCTION}

It is of great interest that machines, specially service robots, interact with humans in a similar manner as a human would. Thus, there is a growing regard to correctly identify the speaker with whom the robot is interacting by their voice alone, and to do so in a real-life setting \cite{Grondin2012,Youssef2010}. In such scenarios, however, new users are often introduced in the environment, such as when a new customer enters a restaurant or when a family member visits the user's home. Accommodating these new users is expected from a service robot, which includes identifying them by their voice. However, typical speaker identification systems require a retraining process every time a new speaker is added \cite{campbell1997}.

In this work, we propose an approach that does not have this requirement (known as \emph{one-shot learning}). It relies on a generic verification model that establishes if two audio recordings are from the same speaker. This is complemented by having an external database of audio recordings of the users to be identified and applying the model to each of its entries to verify if the speech input is of any of the users in the database. Once all the entries are verified, the results are used to establish from which known speaker is the speech input; this process can also deem the user as unknown. Additionally, and more importantly, if the user is unknown and their identification is of interest in the future, it is only required to add their speech input as another entry in the database; the verification model does not requires to be retrained.

The architecture of the proposed model is based on a Siamese Convolutional Neural Network (SCNN). The resulting verification model in this work is able to extract proper audio features and a function that determines the similarity between both inputs so as to verify if the two recordings are from the same speaker. 

This system is planned to be carried out over a service robot in a real-life setting. To benefit the rhythm of the human-robot interaction, the proposed system is expected to perform in a fast manner, and it will be evaluated in this aspect. Additionally, the system is expected to have a high level of performance, so that it does not frequently mistake a user for another and cause frustration. As it is discussed in Section \ref{sec:related_works}, most speaker identification systems in real-life settings have an accuracy of around 80\%, and those that have greater accuracy than this require to know the speakers \emph{a-priori}. Moreover, most service robotics settings assume that there are a limited number of users with which to interact. Thus, we are considering any level of performance above 80\% with a limited number of speakers as acceptable, given the one-shot-learning nature of the proposed system.

A video demonstration of the full system, as well as all relevant downloads (corpora, source code, models, etc.) can be found at \url{http://calebrascon.info/oneshotid/}.

The remainder of this paper is organized as follows: a summary of the related works is presented in Section \ref{sec:related_works}; in Section \ref{sect:model} the proposed system is described, as well as the 3 main components of the core model (training data set, the representation of the data and its architecture); the methodology used for evaluating the models as well as their results are presented in Section \ref{sect:exp}; and, we conclude our work in Section \ref{sect:conclusions}.

\section{RELATED WORKS}\label{sec:related_works}
There is a vast amount of literature on speaker identification, with two important types of techniques: by classification and by verification. The classification type of techniques train one model with an output limited by the classes (or, in this case, known speakers) that it was trained with. When a new speaker enters the scenario, the model attempts to match it with one of the known speakers, resulting in a false positive. On the other hand, the verification type of techniques train a model for each known speaker, with their outputs providing the probability of the speaker being the one with which the model was trained for. This type of techniques are able to register if a speaker is unknown (when all the models provide low probabilities), however, they generally tend to be less accurate than the classification type of techniques \cite{campbell1997}. 

Since the proposed system employs a verifier to measure the similarity between two audio recordings in the speaker domain, the remainder of this section reviews works that are related to this approach.

In \cite{Daqrouq2011} the speaker identification is carried out by comparing a measure of similarity between the audio of a speaker to be identified and the patterns previously generated for the known speakers. Later, in \cite{DAQROUQ2015231}, the author proposes changes to this system where the database and the architecture change, however the similarity measure is kept. Neural networks had also been used with raw audio \cite{Muckenhirn2017}, where a CNN extracts the relevant information and an ad-hoc verifier is generated for every speaker. In \cite{nagrani17}, the authors describe the VoxCeleb database and trained deep learning models for identification and verification of speakers. They use the cosine distance between two signals. In \cite{Mobiny2018}, speaker verification is carried out by using a Siamese model of two Long Short-Term Memory (LSTM) networks, and a contrastive loss for the verification. 

It is important to note that the use of neural networks in conjunction with audio has not been limited only to the identification and verification of the speaker, but has also addressed the issue of extracting audio features to make them more robust. The works of \cite{Snyder2017,Snyder2016,Variani2014,Bhattacharya2017,Heigold2016} use embeddings or complete neural networks to generate features with which statistical methods are applied to verify a speaker.

The aforementioned works, depending on the database used to train, achieved performances above 80\% of accuracy for verification.

Greater performance has been recently achieved, and has been applied in the field of service robotics. For example, in \cite{Grondin2012} a speaker identification process is carried out by inputting the power spectrum of the speech input, a model is trained for each speaker, and then the Euclidean distance between the models output is measured. The authors report a 96\% average rate of identification for 20 speakers. In \cite{Youssef2010}, 10 speakers were identified using 32 MFCC-based characteristics. The reported rate of identification reached 100\% only in certain locations of the speaker, and required such location to be known as part of the training data.

It is very important to mention that in all the aforementioned works that carry out identification via classification, it is assumed that the users to be identified are known. During the testing phase, new speakers are not added, so it is implied that the models are not able to identify speakers that they were not trained for.

\section{PROPOSED SYSTEM}\label{sect:model}

\begin{figure}[thpb]
   \centering
   \framebox{\parbox{3in}{}
   \includegraphics[width=0.4\textwidth]{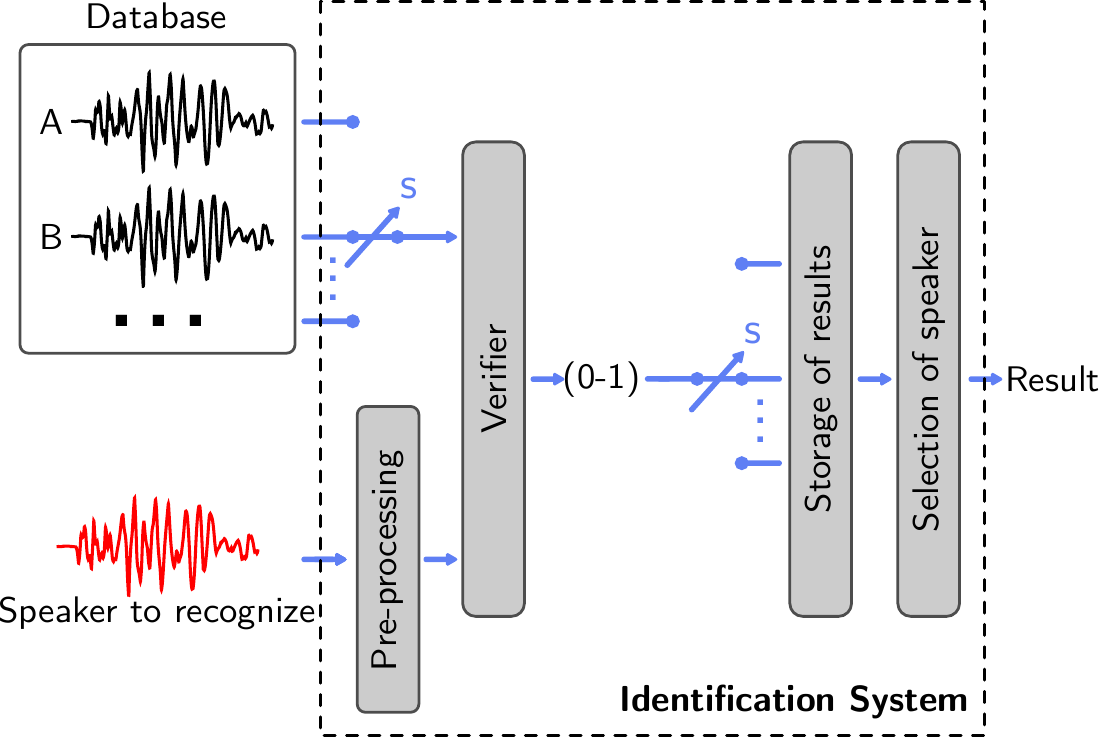}}
   \caption{Diagram for speaker identification.}
   \label{fig:diag_verificador}
\end{figure}

The proposed system is divided into three parts: a generic verifier that outputs the distance between two audio inputs in the speaker domain; an external database that stores audio entries of known users; and a user selector, which has as an input a list of verification scores. A diagram of the complete proposed system is shown in Figure \ref{fig:diag_verificador}.

When an identification is carried out, the system iterates through the database entries, verifying each with the data of the speaker to recognize, and storing each verification result. When the iteration ends, a speaker is selected based on the verification results.

As it can be seen, the central part of the system is the verifier, which is expected to indicate if two audio recordings are from the same user or not. Satisfying this expectation, however, the system provides two virtues that are of great interest to service robotics.

First, the system permits to have several entries per speaker which, as it will be seen, contributes to the robustness of their identification.

\begin{figure}[thpb]
   \centering
   \framebox{\parbox{3in}{}
   \includegraphics[width=0.4\textwidth]{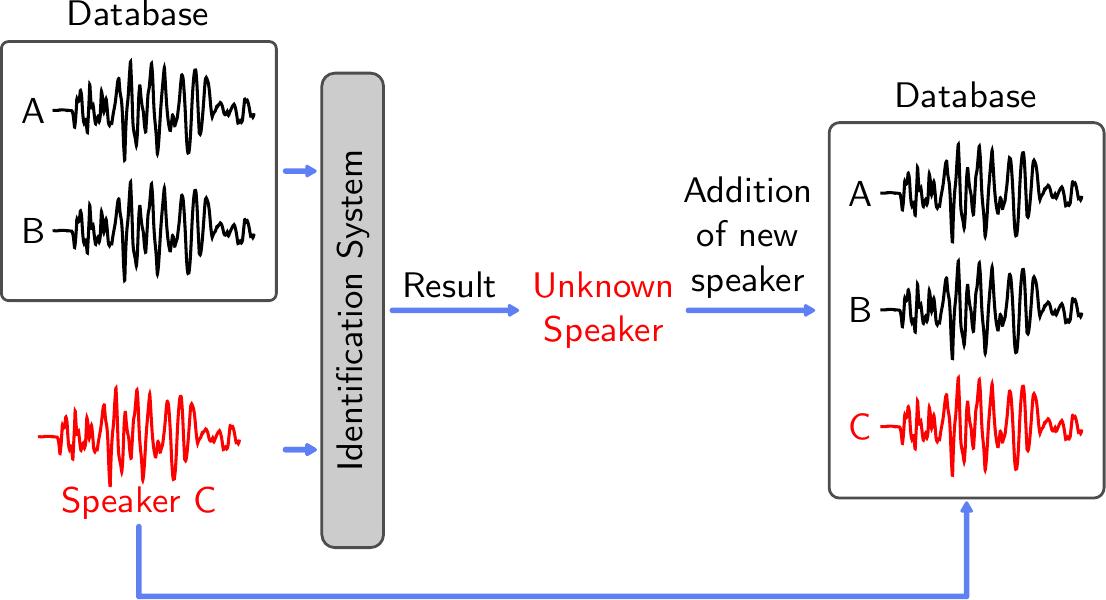}}
   \caption{Diagram for speaker identification with speaker addition.}
   \label{fig:diag_ingreso}
\end{figure}

Second, the system also permits the addition of new speakers without requiring to retrain the model. When the selection process deems the new audio data as not belonging to any of the known speakers, a human-robot interaction can be carried out to ask the unknown user their name and add them to the database for future identifications. This operation is summarized in Figure \ref{fig:diag_ingreso}.

As it can be concluded, it is very important that the generic verifier at the center of the system is trained with a database with a large number of speakers. This is so that it can attain its generalizability. Additionally, it is also important that the audio data is appropriately represented and that a proper architecture is chosen, so the system as a whole achieves the performance and speed previously discussed. These three aspects (databases, data representation, and architecture) are described in the remainder of this section.

\subsection{Databases}
\label{sect:databases}

The LibriSpeech database \cite{vassil2015librispeech} is based on the project LibriVox. It was recorded under a controlled environment, with just one speaker talking at a time, with little variance of background noise between recordings. This database was chosen since it is text-independent, which will make the model robust against any phrasing. It contains more than 100 speakers, which will grant the model generality in its verification. And it does not involve any monetary requirement, which simplified its procurement.

Another database that was used is Voxceleb \cite{nagrani17}. It was obtained from YouTube videos of interviews of celebrities. This resulted in the database having more than 6000 speakers recorded in many different recording environments, which varied in terms of noise presence and distance of the microphone to the user. This database was chosen because of these variations, since it is expected to provide the verification model robustness against them. Additionally, it also does not involve any monetary requirement, which simplified its procurement.

For training, 80\% of the databases was used, with 10\% for validation and 10\% for testing. Since the focus of the proposed approach is to verify generically between speakers, the training, validation and testing datasets do not share speakers. Meaning, none of the recorded data of the speakers of the validation and testing datasets was used as part of the training process.

For further testing, an evaluation corpus was recorded based on the testing subset of LibriSpeech, referred to here as LibriSpeechReal. Two real environments were used: an open-cubicle computer lab (background noise at -45 dBFS with a $\tau_{60} \approx 0.51 s$) and an office (background noise at -47 dBFS with a $\tau_{60} \approx 0.39 s$). A monitor speaker reproducing the Librispeech testing subset data was placed 1 m. from a flat-response microphone, which recorded such reproduction. This corpus emulates what a service robot would be hearing (in terms of noise and reverberation) from a speaker in a real-life setting.

\subsection{Representation of the data}\label{subsec:rep_dat}

The audio signals are converted to a time-frequency spectrogram before being fed to the model. Two variations of this spectrogram-based data representation were used as inputs for the trained models. Both variations are generated using a 1 s. segment of audio, using an overlap of 50\%, and only using frequencies of the lower half of the frequency spectrum as the input to the model. Preliminary testing showed no major differences in performance between using only the lower half of the spectrum and using the full frequency range. Both variations are normalized as part of their calculation.

One variation, referred to here as Spect 256, applies a 1024 FFT point window, and only uses the lower 256 FFT points of the frequency range. These 256 FFT points represent frequencies up to 4 $kHz$.

The other variation, referred to here as Spect 32, applies a 400 FFT point window, and only uses the lower 32 FFT points of the frequency range. These 32 FFT points represent frequencies up to 1.28 $kHz$.

It is important to note that other types of representation were tested, such as the FFT of the whole audio segment, the Mel-Frequency Cepstral spectrum of the whole audio segment, as well using the upper half of the frequency range. However, the performance obtained with these representations did not improved upon the results when using the representations described in this section.

\subsection{Architecture}\label{subsec:arch}

Several architectures were tested to be used as a generic verifier. In this section, the three models that obtained the highest performance are described. These three models are based on two architectures: VGG \cite{simonyan2014very} and ResNet 50 \cite{he2015resnet}. 

In the proposed models, these architectures are arranged as a Siamese network \cite{Bromley1993,LeCunsiames2006}. When used for verification, these networks consists mainly of two elements: feature extraction and similarity calculation. To this effect, their first layers are mainly convolutional layers, while the last layers are fully connected. The outputs of the latter are passed through a SoftMax function to calculate the probability of the two inputs being the same.

For each model, 800,000 audio files were randomly selected from the used  database (either Librispeech or Voxceleb) for training per epoch, while 80,000 data files were randomly selected for validation, and 80,000 for testing. Each data set had the same amount of positive and negative examples. Although the audio files are the same through the epochs, the audio input that is fed to the models is different due to the random selection of the segment of audio from the file.

\subsubsection{VGG}

\begin{figure}[thpb]
   \centering
   \framebox{\parbox{3in}{}
   \includegraphics[width=0.3\textwidth]{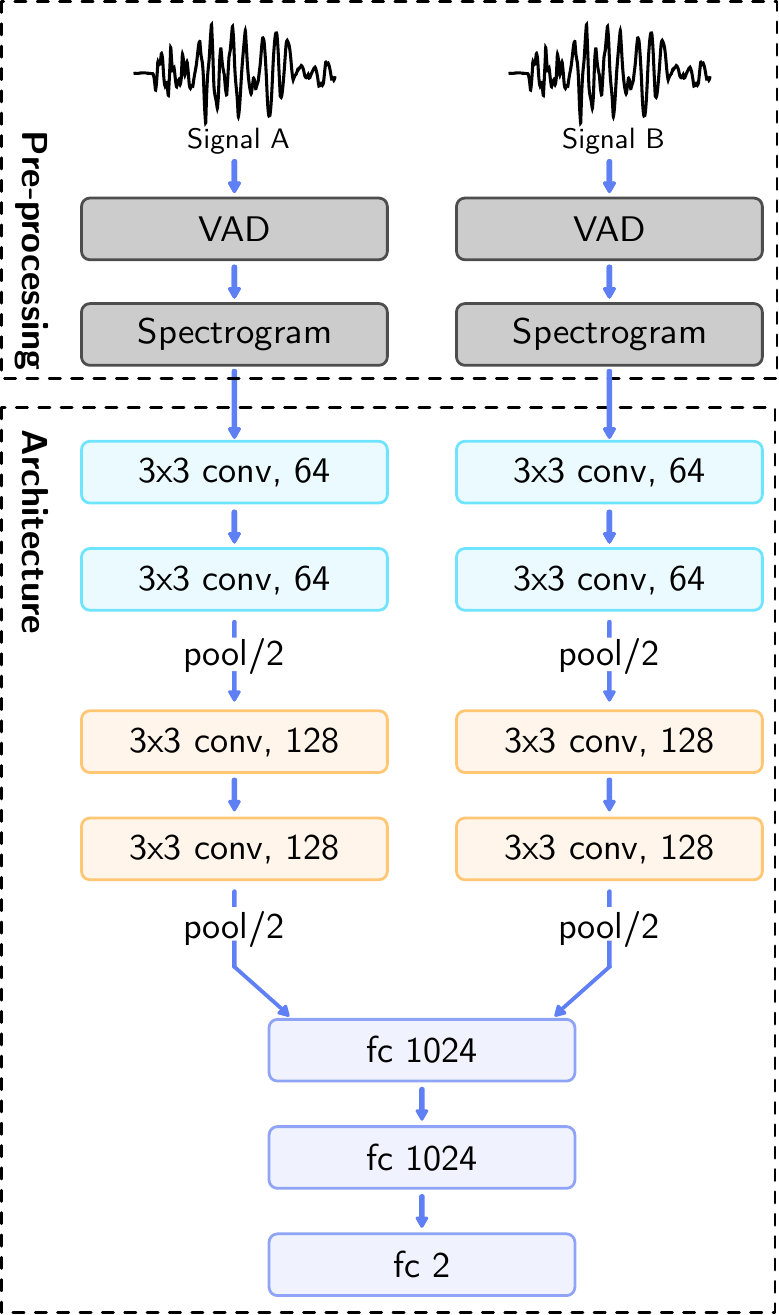}}
   \caption{Siamese VGG7.}
   \label{fig:modulo_vgg}
\end{figure}

In Figure \ref{fig:modulo_vgg}, the proposed Siamese network inspired by the VGG 16 \cite{simonyan2014very} can be seen. It is composed of: 4 convolutional layers for each part of the Siamese network; 2 pooling layers; and 3 fully-connected layers for identification. This network corresponds to the first 4 layers and the last 3 layers of the original VGG 16 architecture. Since this network only use 7 layers of the VGG 16 architecture, it is referred to here as VGG7.

The convolutional layers use 64 filters of $3\times3$ for the first two layers and 128 filters for layers 3 and 4. The intermediate layers use ReLU as activation function.

Two of the proposed models are based on this network, each using the different data representation discussed earlier. The model referred to here as VGG7$_{256}$ uses the the Spect 256 data representation, while the model referred to here as VGG7$_{32}$ uses the Spect 32 data representation.

The training of this network was done in batches of 50 samples, using cross entropy as loss and stochastic gradient descent as optimization algorithm. This model at first was trained for 15 epochs with a learning rate of 0.01 with Librispeech; another model was later trained for 8 epochs with the same learning rate with VoxCeleb. 

\subsubsection{ResNet 50}\label{subsubsec:resnet}

\begin{figure}[thpb]
   \centering
   \framebox{\parbox{3in}{}
   \includegraphics[width=0.3\textwidth]{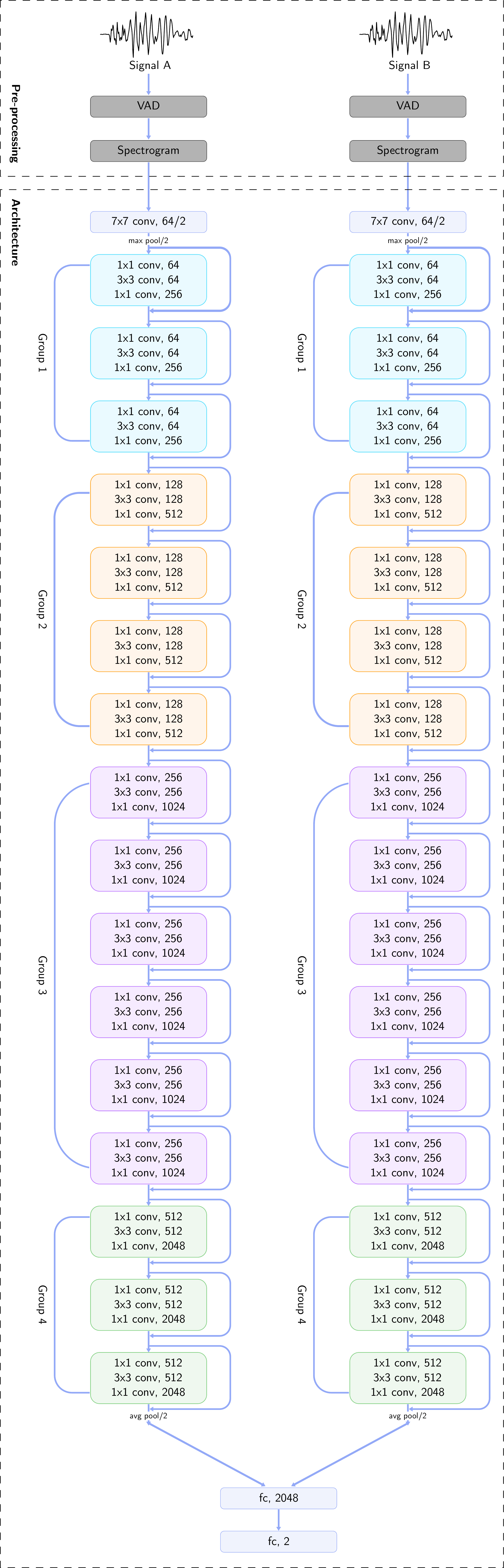}}
   \caption{Siamese ResNet 50.}
   \label{fig:resnet_50_siames}
\end{figure}

In Figure \ref{fig:resnet_50_siames}, the proposed Siamese network based on the ResNet 50 \cite{he2015resnet} can be seen. It is composed of: 1 convolutional layer with 64 filters of $7\times7$; 16 bottleneck blocks; and 2 fully connected layers. Intermediate layers use batch normalization and ReLU as activation function.

The proposed model based on this network is referred to here only as ResNet for ease of reference, and it uses the Spect 32 data representation.

To speed up the training process, an initialization network similar to ResNet was used, with the last layer performing both speaker verification and speaker classification. After an acceptable loss was achieved, the initialized weights were re-used to start the training of ResNet.

The training of this network was done with Librispeech in batches of 10 samples, using cross entropy as loss with l2 norm as regularizer and stochastic gradient descent as optimization algorithm. This model, after the initialization, was trained for 10 epochs with a learning rate of 0.01, and later another 10 epochs with a learning rate of 0.001.

\section{EVALUATION AND RESULTS} \label{sect:exp}

The three models previously described were evaluated as part of the proposed system that can add new speakers to identify. The models were evaluated in four aspects: the performance of the verifier, the performance of the system as a classifier using clean audio, the speed of the system, and the accuracy of the system when it is evaluated in real-life settings. As mentioned before, an accuracy higher than 80\% is considered acceptable, since the closest system to our proposed approach is \cite{nagrani17}, and it had an accuracy of 80.5 \%.

\subsection{Evaluation of verifier}

\begin{figure}[thpb]
   \centering
   \framebox{\parbox{3in}{}
   \includegraphics[width=0.45\textwidth]{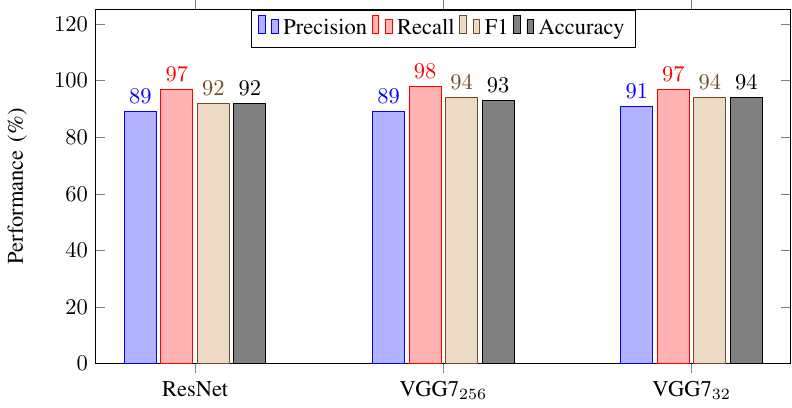}}
   \caption{Evaluation of verifier with 1000 audios.}
   \label{fig:eval_ver}
\end{figure}

The different trained verifiers were evaluated with 1,000 samples, 500 samples from the same speakers and 500 from different speakers. These audios were randomly selected from the testing data set. This process is performed 10 times by each verifier and the average of true positives, true negatives, false positives and false negatives of these results are obtained. Then, the precision, recall, F1 and accuracy are calculated, and are shown in Figure \ref{fig:eval_ver}.

As it can be seen, the model with the highest precision is the VGG7$_{32}$, reaching 91\%, followed by VGG7$_{256}$ and ResNet with 89\%. Although the models have different results in the different metrics, they do not differ in more than 2\%. All results are above the desired 80\% which implies that the proposed verification models function appropriately for comparing whether two audio signals are from the same speaker or not. It is important to mention that these results only evaluate the performance of the models as standalone generic verifiers, and do not use an external database to carry out the identification process.

\subsection{Evaluation as a classifier} \label{subsec:ver_as_clas}

To evaluate the system as a classifier, the external database is considered as static a-priori knowledge of known users. The final selection is the speaker with the highest average verification result of all the users in the database against the speaker to recognize. 

\begin{figure}[thpb]
   \centering
   \framebox{\parbox{3in}{}
   \includegraphics[width=0.42\textwidth]{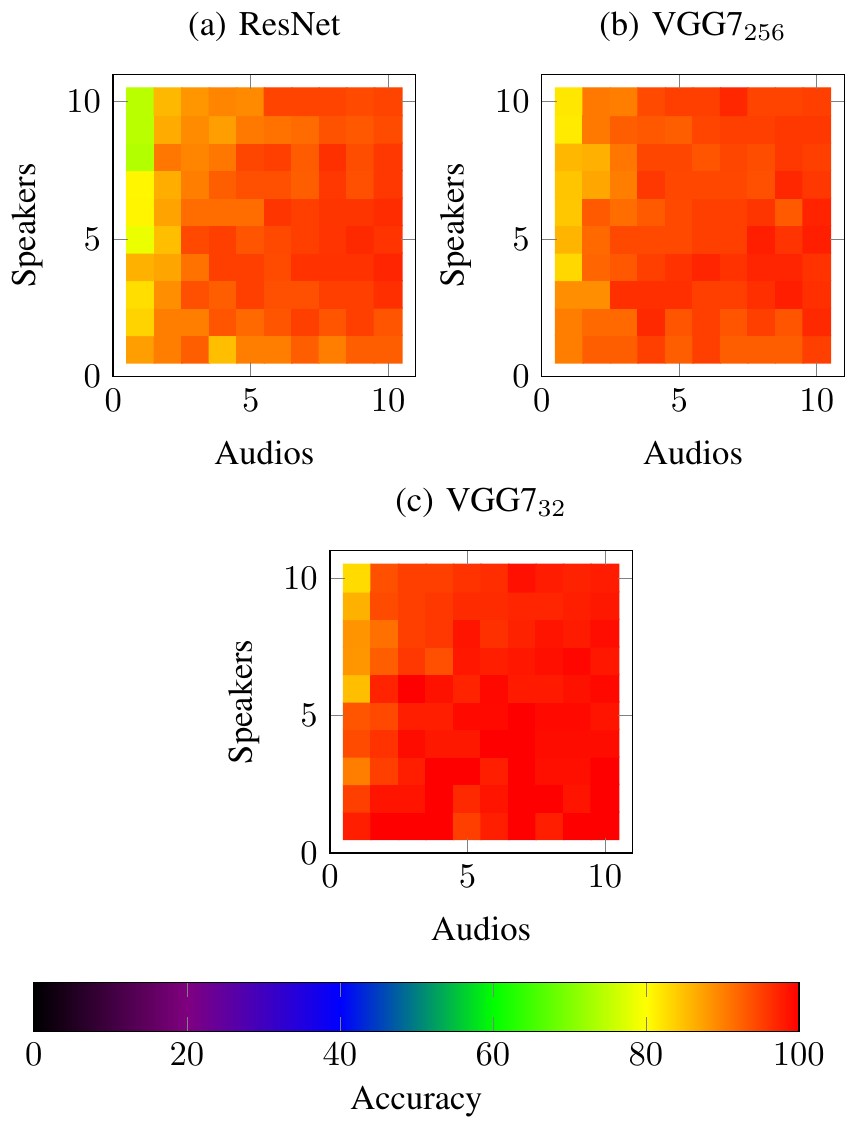}}
   \caption{Accuracy heat maps as classifier. `Audios' indicates the amount of audios known per speaker. `Speakers' indicates the number of known speakers.}
   \label{fig:mapa_calor_clas_librispeech}
\end{figure}

In this evaluation, an accuracy heat map is obtained for each model. Each heat map presents the accuracy of the system having different combinations of number of known speakers and number of audio recordings the database has of each known speaker. Each speaker-audios combination is tested such that each known speaker is verified against known speakers and unknown speakers alike in a balanced manner. The intent of the accuracy heat maps is to show how the performance of the system changes as it has more known users and more audio recordings per known user. These accuracy heat maps are shown in Figure \ref{fig:mapa_calor_clas_librispeech}.

As it can be seen, the three models have a similar pattern: as the number of speakers increases, more known audio recordings in the external database are required to achieve a better performance.

However, it can be reported that the three systems obtained an accuracy higher than 70\% in all speaker-audios combinations, while some combinations reached values very close to 100\%. Additionally, VGG7$_{32}$ is the model with the best classifying results with an accuracy of 97.7\%, followed by 95\% of the VGG7$_{256}$ and 94.5\% of the ResNet. Additionally, the overall average accuracy of the system with VGG7$_{32}$ model is of 97\%, which implies that it performed with a high accuracy in all speaker-audios combinations.

\subsection{System speed}\label{subsec:model_ef}

\begin{figure}[thpb]
   \centering
   \framebox{\parbox{3in}{}
   \includegraphics[width=0.45\textwidth]{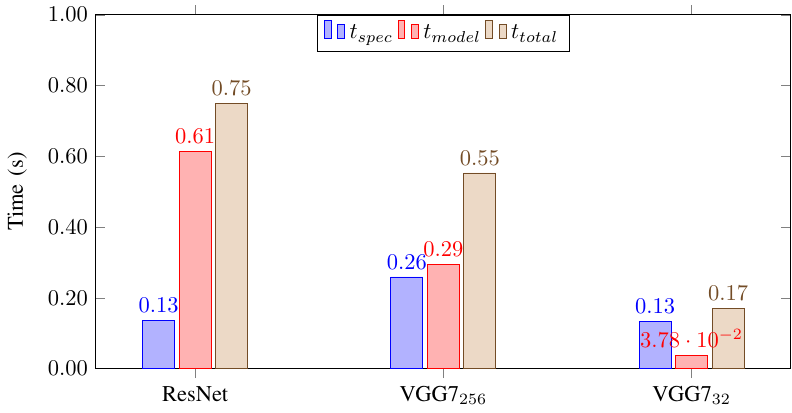}}
	\caption{Average run time for each model. $t_{spec}$ indicates the run time of the calculation of the frequency spectrogram of input data and loading the spectrogram database. $t_{model}$ indicates the run time of the verification process to run the CNN model through all the database entries. And $t_{total}$ indicates the total run time.}	
	\label{fig:duracion}
\end{figure}

To determine the speed the system when using each model, the average run time was measured of 10 sets of verifications of 1 audio segment against 100 others that are stored in the database as time-frequency spectrograms generated from 1-second audio segments. This was carried out to avoid calculating the spectrograms of the audio data stored in the database for each verification. The times measured for each verification were: the time to calculate the spectrogram of the input audio and load the spectrogram database ($t_{spec}$), and the time that the verification process takes to run the CNN model through all the database entries ($t_{model}$). These run times are shown in Figure \ref{fig:duracion}.

As it can be seen in Figure \ref{fig:duracion}, the VGG-based models are faster than ResNet, which is to be expected since they are smaller and require less computations. VGG7$_{32}$ is much faster than VGG7$_{256}$, in part because VGG7$_{256}$ architecture has much more parameters, and because calculating a 1024 FFT point spectrogram takes more time than calculating a 400 FFT point spectrogram.

It is important to note that in the VGG-based models, the calculation of the spectrogram takes up a considerable amount of their total run time. In the case of VGG7$_{32}$, it takes 76.5\% of the total run time; in the case of the VGG7$_{256}$, it takes 47.3\%. This points that, if the system is to be made faster, it may be necessary to find a faster way to calculate the spectrogram, or find another type of data representation that is faster to calculate. However, it is to be noted that with the VGG7$_{32}$ model, the total run time of the system is 0.17 s. which is an acceptable response time given its one-shot-learning nature. In addition, the bulk of the total run time is taken by the calculation of the spectrogram of the input data and the loading of the database, both of which only need to be carried out once. This indicates that the response time will not significantly increase as the number of users increases; not unless it reaches several order of magnitudes bigger than 10 users, which is unlikely given the application case of service robotics.

\subsection{Evaluation in real-life settings}\label{subsec:real_env}

\begin{figure}[thpb]
   \centering
   \framebox{\parbox{3in}{}
   \includegraphics[width=0.45\textwidth]{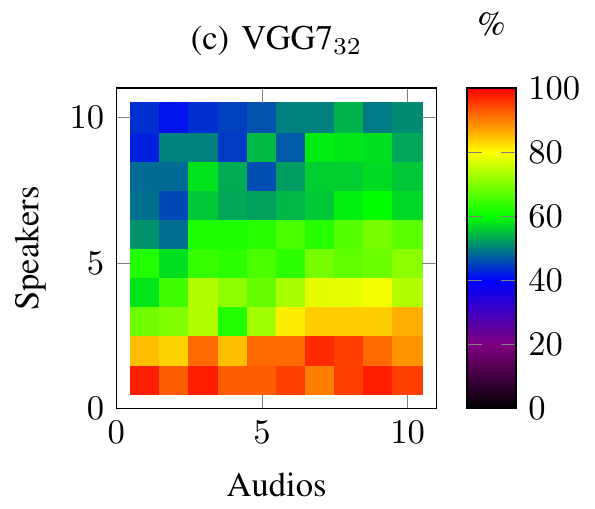}}
   \caption{Accuracy heat maps with LibriSpeechReal for a model trained with LibriSpeech dataset. `Audios' indicates the amount of audios known per speaker. `Speakers' indicates the number of known speakers.}
   \label{fig:mapa_calor_clas_libriaira_t_librispeech}
\end{figure}

Up until this point, the evaluated models were trained with LibriSpeech, which is a clean data set. However, given the service robotics application case, it is of interest to evaluate the system in a real-life setting. For this purpose, an evaluation corpus based on LibriSpeech was recorded in two real environments. This corpus is refereed to here as LibriSpeechReal (details are provided in Section \ref{sect:databases}). The VGG7$_{32}$ model is chosen due to it being the most accurate model as well as being the fastest, (as seen in Sections \ref{subsec:ver_as_clas} and \ref{subsec:model_ef}), the two desired aforementioned qualities relevant for the service robotics application case. The resulting accuracy heat map is shown in Figure \ref{fig:mapa_calor_clas_libriaira_t_librispeech}.

As it can be seen, the performance of the system decreases considerably, up to 51\% in many cases. This is due to the lack of robustness of the verification model against noise and reverberation, since it was trained with a clean dataset.

\begin{figure}[thpb]
   \centering
   \framebox{\parbox{3in}{}
   \includegraphics[width=0.45\textwidth]{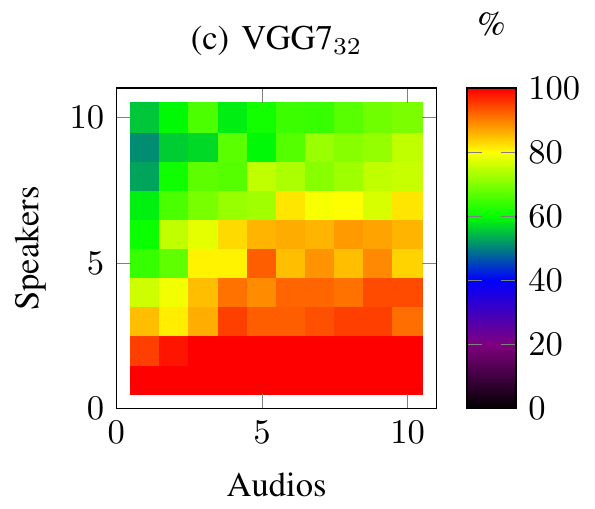}}
   \caption{Accuracy heat maps with LibriSpeechReal for a model trained with VoxCeleb dataset. `Audios' indicates the amount of audios known per speaker. `Speakers' indicates the number of known speakers.}
   \label{fig:mapa_calor_clas_libriaira_t_voxceleb}
\end{figure}

To overcome this situation, the VGG7$_{32}$ model was retrained with the VoxCeleb database \cite{nagrani17} (described in Section \ref{sect:databases}). In this case, more than 2.5 million samples were used per training epoch. The evaluation with LibriSpeechReal of the retrained VGG7$_{32}$ model is shown as a heat map in Figure
\ref{fig:mapa_calor_clas_libriaira_t_voxceleb}.

As it can be seen, its performance has increased significantly, now obtaining an average accuracy of 81.2\%. More importantly, it can also be seen that the system overcomes the accuracy threshold expected for a service robot in many of the speaker-audios combinations.

For further analysis, two confusion matrices are shown in Tables \ref{table:matrix_10_1} and \ref{table:matrix_10_10}, to allow us to analyze in which speakers is the model failing. The matrices belong to the combination of 10 known speakers with 1 audio per speaker, and to the combination of 10 known speakers with 10 audios per speaker. These two speaker-audios combinations were chosen since they present the most wide variety of change when increasing the amount of audios per speaker. It's worth mentioning that the confusion matrices for the speaker-audios combination for 1 or 2 known speakers, independently of the amount of audios, score a near perfect result, with the model not confusing any of the known speakers or the unknown ones.

\begin{table}[H]
\centering

\begin{tabular}{|p{0.02\textwidth}|p{0.012\textwidth}|p{0.012\textwidth}|p{0.012\textwidth}|p{0.012\textwidth}|p{0.012\textwidth}|p{0.012\textwidth}|p{0.012\textwidth}|p{0.012\textwidth}|p{0.012\textwidth}|p{0.012\textwidth}|p{0.012\textwidth}|}

\cline{2-11}
\multicolumn{1}{c}{} & \multicolumn{10}{|c| }{\bfseries Known speakers} \\
\hline

\bfseries Sp. & \bfseries 1 & \bfseries 2 & \bfseries 3 & \bfseries 4 & \bfseries 5 & \bfseries 6 & \bfseries 7 & \bfseries 8 & \bfseries 9 & \bfseries 10 & \bfseries U \\ 
\hline
\bfseries 1 & 16 & 1 & 3 & 0 & 0 & 0 & 0 & 0 & 0 & 0 & 0 \\ 
\hline
\bfseries 2 & 0 & 10 & 0 & 1 & 0 & 0 & 2 & 1 & 0 & 6 & 0 \\ 
\hline
\bfseries 3 & 7 & 0 & 4 & 2 & 0 & 0 & 0 & 0 & 6 & 1 & 0 \\ 
\hline
\bfseries 4 & 3 & 3 & 0 & 3 & 0 & 7 & 1 & 0 & 3 & 0 & 0 \\ 
\hline
\bfseries 5 & 1 & 0 & 0 & 2 & 9 & 0 & 5 & 0 & 0 & 3 & 0 \\ 
\hline
\bfseries 6 & 1 & 0 & 0 & 0 & 0 & 19 & 0 & 0 & 0 & 0 & 0 \\ 
\hline
\bfseries 7 & 0 & 1 & 0 & 2 & 5 & 0 & 6 & 3 & 0 & 3 & 0 \\ 
\hline
\bfseries 8 & 0 & 0 & 0 & 2 & 0 & 2 & 1 & 9 & 0 & 6 & 0 \\ 
\hline
\bfseries 9 & 1 & 2 & 2 & 1 & 0 & 0 & 0 & 0 & 14 & 0 & 0 \\ 
\hline
\bfseries 10 & 2 & 1 & 1 & 0 & 0 & 0 & 1 & 1 & 0 & 14 & 0 \\ 
\hline
\bfseries U & 0 & 0 & 0 & 0 & 0 & 0 & 0 & 1 & 0 & 1 & 18 \\ 
\hline

\end{tabular}
\caption{Confusion matrix for the combination of 10 known speakers with 1 spectrogram in the database. `Sp.' represents the label of the speaker and `U' represent unknown speakers.}
\label{table:matrix_10_1}
\end{table}

\begin{table}[H]
\centering

\begin{tabular}{|p{0.02\textwidth}|p{0.012\textwidth}|p{0.012\textwidth}|p{0.012\textwidth}|p{0.012\textwidth}|p{0.012\textwidth}|p{0.012\textwidth}|p{0.012\textwidth}|p{0.012\textwidth}|p{0.012\textwidth}|p{0.012\textwidth}|p{0.012\textwidth}|}

\cline{2-11}
\multicolumn{1}{c}{} & \multicolumn{10}{|c| }{\bfseries Known speakers} \\
\hline

\bfseries Sp. & \bfseries 1 & \bfseries 2 & \bfseries 3 & \bfseries 4 & \bfseries 5 & \bfseries 6 & \bfseries 7 & \bfseries 8 & \bfseries 9 & \bfseries 10 & \bfseries U \\ 
\hline
\bfseries 1 & \cellcolor{green!25} 20 & 0 & 0 & 0 & 0 & 0 & 0 & 0 & 0 & 0 & 0 \\ 
\hline
\bfseries 2 & 0 & \cellcolor{green!25} 13 & 0 & 0 & 0 & 0 & 0 & 3 & 0 & 4 & 0 \\ 
\hline
\bfseries 3 & 4 & 0 & \cellcolor{green!25} 12 & 1 & 0 & 0 & 0 & 0 & 3 & 0 & 0 \\ 
\hline
\bfseries 4 & 0 & 0 & 0 & \cellcolor{green!25} 15 & 0 & 4 & 0 & 0 & 1 & 0 & 0 \\ 
\hline
\bfseries 5 & 0 & 3 & 0 & 0 & \cellcolor{red!25} 7 & 0 & 7 & 0 & 0 & 3 & 0 \\ 
\hline
\bfseries 6 & 0 & 0 & 0 & 0 & 0 & \cellcolor{green!25} 20 & 0 & 0 & 0 & 0 & 0 \\ 
\hline
\bfseries 7 & 0 & 0 & 0 & 0 & 2 & 0 & \cellcolor{green!25} 15 & 0 & 0 & 3 & 0 \\ 
\hline
\bfseries 8 & 0 & 1 & 0 & 0 & 0 & 0 & 3 & \cellcolor{green!25} 10 & 0 & 6 & 0 \\ 
\hline
\bfseries 9 & 1 & 0 & 2 & 7 & 0 & 1 & 0 & 0 & \cellcolor{red!25} 9 & 0 & 0 \\ 
\hline
\bfseries 10 & 0 & 1 & 0 & 0 & 0 & 0 & 1 & 4 & 0 & \cellcolor{blue!25} 14 & 0 \\ 
\hline
\bfseries U & 0 & 0 & 0 & 0 & 0 & 0 & 2 & 0 & 0 & 0 & \cellcolor{blue!25} 18 \\ 
\hline

\end{tabular}
\caption{Confusion matrix for the combination of 10 known speakers with 10 spectrograms in the database. `Sp.' represents the label of the speaker and `U' represent unknown speakers. A green cell represents an improvement over Table \ref{table:matrix_10_1}, a blue cell represents no improvement, and a red cell represent a decline.}
\label{table:matrix_10_10}
\end{table}

As it can be seen, there is a considerable improvement in most of the users in the confusion matrix shown in Table \ref{table:matrix_10_10}, which implies that having more audio entries per speaker contributes to its performance. However, this change does not seem to affect the detection of unknown speakers, which is to be expected, since audio entries of unknown speakers are not stored in the database.

It can also be seen that there is one user that was not affected by the change in number of audio entries and that there are two users that were mistaken more with more audio entries. This is contrary to the tendency shown by the majority of the users. Although we are not able to explain definitively why this happened, we believe that it may be attributed to the nature in which the audio entry is selected: the system asks the user to talk for a pre-specified amount of time; then a 1 s. audio segment is randomly chosen from the recording that has an average energy above a pre-specified threshold. This was carried out as a form of Voice Activity Detection (VAD) to ensure that the audio segment stored in the database had speech information with which to identify the user in the future. However, using this basic form of VAD may result in storing segments of audio with a considerable amount of silence in them. When additional testing was carried out, it was found that modifying the VAD threshold did not impact in any meaningful way the overall performance of the system. Thus, a more sophisticated VAD system may be required. Moreover, other data representation techniques could be used to normalize the input in terms of energy as well as feature extraction.

\section{CONCLUSIONS} \label{sect:conclusions}

In this work, a speaker identification system was proposed based on a generic verifier and a dynamic external database of audio recordings of known speakers. Because of its generality, this system does not need to be retrained when new speakers enter the scenario, since these can be flexibly added to the database. For a demonstration and access to all relevant information, visit \url{http://calebrascon.info/oneshotid/}.

The performance of the highest three tested models for the role of the generic verifier were shown. These were Siamese convolutional models, based on two architectures: VGG 16 and ResNet 50. 

The overall performance of the system showed an average accuracy of 97\% with a clean testing corpus and an average performance of 81.2\% with real-life recordings. However, a compromise needs to be struck between the amount of audio entries per known speaker stored in the database and the number of speakers that it needs to identify.

Speed evaluations showed that the VGG7$_{32}$ is the fastest model of the proposed ones, and its response to verify 1 audio against 100 hundred spectrograms is 0.17 s. This is an acceptable response time, given the one-shot-learning nature of the system as a whole.

A model trained with a noisy database, proved to be robust against noise, and achieve an accuracy above the desired 80\% in a real-life setting with most combinations of number of speakers vs audios entries per speaker. Interestingly, most of these combinations are aligned with the application case of service robotics.

It is left for future work to integrate this system as part of a task of a service robot and to test its in the robot's social interaction. Additionally, the training parameters of the proposed models could be refined to further improve the verification performance. Finally, as mentioned before, a more sophisticated VAD system could be employed and other types of data representation could be explored.

\section*{ACKNOWLEDGMENT}

The authors would like to thank the support of CONACYT through the research grant 251319, UC-MEXUS through the research grant CN-17-54, and PAPIIT-UNAM through the research grant IA104016.

\bibliographystyle{IEEEtran}
\bibliography{myrefs}

\end{document}